\documentclass[twocolumn,showkeywords,preprintnumbers,amsmath,amssymb]{revtex4}
\usepackage{graphicx}% Include figure files
\usepackage{dcolumn}% Align table columns on decimal point
\usepackage{bm}% bold math
\usepackage{enumitem} 
\begin{document}

\title{Dark Energy and Consciousness}% Force line breaks with \\

\author{Daegene Song}

\affiliation{%
Department of Management Information Systems, Chungbuk National University, Cheongju, Chungbuk 28644, Korea
}%

\date{\today}% It is always \today, today,

\begin{abstract}

One of the most important concepts in logic and the foundations of mathematics may be useful in providing an explanation for the cosmological constant problem. A connection between self-reference and consciousness has been previously discussed due to their similar nature of making a reference to itself. Vacuum observation has the property of self-reference and consciousness in the sense that the observer is observing one's own reference frame of energy. In this paper, the cyclical loop model of self-reference is applied to the vacuum observation, such that the discrepancy between the energy density resulting from the first part of the causal loop (i.e., the classical irreversible computation of the observer's reference frame) and the other part of the causal loop (i.e., nondeterministic quantum evolution) corresponds to $\sim 10^{123}$. This effectively provides a consistent explanation of the difference between the observed and the theoretical values of the vacuum energy, namely, the cosmological constant problem.

\end{abstract}

\maketitle
\section{introduction}
Unlike the commonly held perception, a vacuum is surprisingly strange and mysterious. Although the name implies emptiness or nothingness, the reality is quite different. Before the twentieth century, to come up with an explanation of the mysterious nature of light, a vacuum was thought to be filled with aether that acted as a medium for light waves. After Einstein formulated the theory of special relativity, the idea of an aether-filled vacuum faded away. However, with the development of quantum theory, it soon became evident that even vacuums contain non-zero energy. This quantum effect again became problematic with physics at the cosmological scale. While the cosmological constant from observation yielded a value close to zero, the theoretical computation yielded a huge discrepancy  \cite{carroll,CC1,CC2,ryden}.

On the other hand, the concept of self-reference - an object that refers to itself - has been one of the central themes used in the study of logic \cite{self}. The simplest well-known example of self-reference is the so-called the Liar's Paradox which may be written as follows: 
\begin{equation}
This \; statement \; is\; not\; true 
\label{liar}\end{equation}
If we assume the sentence is true, then it says it is false. When the sentence is assumed to be false then it must be true. Therefore, either way there is a contradiction. The logical structure seen in  (\ref{liar}) can be found in various problems, such as Russell's paradox in set theory \cite{russell} and G\"odel’s incompleteness theorems in mathematical logic \cite{godel} (see \cite{lucas} for a review). On the other hand, understanding consciousness has been studied extensively by neuroscience \cite{koch,dahaene} and in the neural network field  \cite{harvey}. Understanding consciousness, however, has been a formidable task because of its unique property of self-reference. Unprecedented in other physical systems, this strange and singular ability to consciously refer to self bears a resemblance to the self-reference of logic.

\begin{figure}
\begin{center}
{\includegraphics[scale=.9]{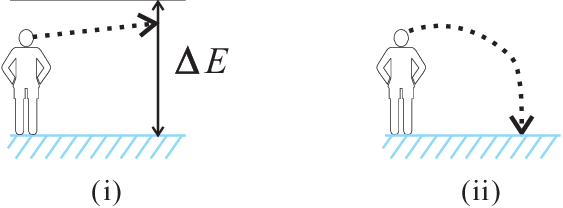}}

\end{center}
\caption{(i) Ordinarily, the observer observes with respect to the reference frame, which therefore leads to the observation of the energy difference as seen in quantum theory. (ii) However, in the case of self-observing consciousness, the observer observes his or her own reference frame. This is vacuum observation.  }
\label{Va}\end{figure}

%%%%%%%%%%%%%%%%%%

How could the problem of discrepancy in the cosmological constant be related to self-reference as in (\ref{liar})?   As discussed in  \cite{song2009}, the degrees of freedom for the reference frame ought to be the same as that of the state vectors. For example, if the outcome of a spin is up or down, the reference frame in measuring the spin is up or down as well. Comparably, regarding the observation of energy states, the observer's reference frame is in one of the states as well; therefore, the observer observes only the relative difference (Fig. \ref{Va} (i)). However, in the case of a vacuum, the observer is observing his or her own reference frame. For energy observation, the vacuum corresponds with the observer observing his or her own reference frame of the ground state energy (Fig. \ref{Va} (ii)).

This paper will consider consciousness in terms of a cyclical loop used in self-reference, particularly regarding time, and attempt to explain the discrepancy between the theoretical and observational values of vacuum energy. The cosmological constant problem is reviewed in Section 2, and the cyclical loop model of self-reference and consciousness is discussed in Section 3. The analyzed case is then applied to the computational model in Sections 4 and 5. We then conclude with a few brief remarks.

%%%%%%%%%%

\section{Review}
When the general theory of relativity was first published in 1915, the universe was not known to be expanding or shrinking, at least on a cosmological scale. Indeed, to come up with a static model of the universe, Einstein originally introduced an additional constant term, $\Lambda$, in his field equation. Based on the assumption that the universe is isotropic, meaning it looks symmetric rotationally, and homogeneous, meaning its density is roughly the same anywhere, the metric can be greatly simplified on a sufficiently large scale. From the metric, Einstein's field equation yields a solution found by Friedman.

The goal then was to see if the experimental observation approximated the actual value of the cosmological constant, $\Lambda$. In 1998, two research teams \cite{Observation1,Observation2}  measured the cosmological parameters by observing a Type Ia supernova. Surprisingly, the researchers found that the universe was accelerating in its expansion with a non-zero cosmological constant value. To determine the value of $\Lambda$, astronomers used an equation that relates the luminosity of distant stars and the redshift, which can show the cosmological constant value to be
\begin{equation}
\Lambda \sim 10^{-35}{\rm{s}}^{-2}
\label{lambda}\end{equation} 
while the vacuum energy density, $\rho_{\Lambda}$, is estimated to be 
\begin{equation}
 \frac{\Lambda c^2}{8\pi G} \sim 5.35 \times 10^{-10} J\cdot m^{-3} 
\label{rho}\end{equation}
where 
\begin{equation}
G\approx 6.7 \times 10^{-11} m^3 kg^{-1} s^{-2}
\end{equation} 
is the Newton's gravitational constant.

To compare it with theoretical value, we may consider the energy density predicted from quantum theory, specifically, Planck energy density, $\rho_{p}$, which is arrived at by dividing the Planck mass 
\begin{equation}
m_{p} = \sqrt{\frac{\hbar c}{G}} \sim 2.17 \times 10^{-8} kg
\end{equation}
by the volume obtained from the cube of the Planck length, 
\begin{equation}
l_{p} = \sqrt{\frac{\hbar G}{c^3}} \sim 1.61 \times 10^{-35} m,
\end{equation} 
which then yields 
\begin{equation}
\frac{m_{p} c^2}{l_{p}^3} \sim 10^{113} J\cdot m^{-3}
\label{quant}\end{equation}
Comparing the theoretical value in (\ref{quant}) with the observational one in (\ref{rho}) yields the following: 
\begin{equation}
\frac{\rho_{p}}{\rho_{\Lambda}} \sim 10^{123}
\label{discre}\end{equation}
The discrepancy in (\ref{discre}) is one of the important problems in physics. There have been several suggestions for resolving this enormous difference between the theoretical prediction and the experimental outcome; however, the problem is generally considered to be unsolved \cite{carroll,CC1,CC2}.

\begin{figure}
\begin{center}
{\includegraphics[scale=.55]{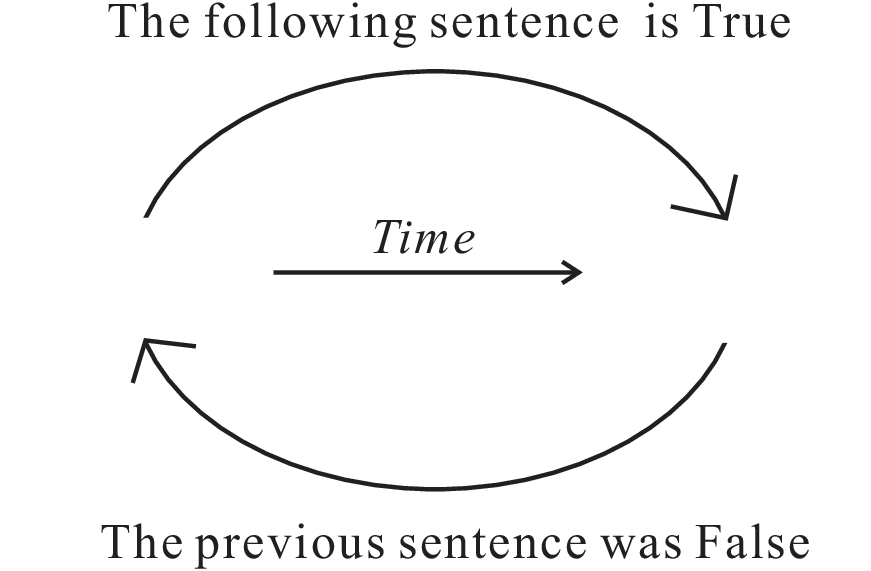}}

\end{center}
\caption{ Cyclical loop of the Liar's Paradox. First the statement refers to a future event in a time-forward manner. Next, the looping occurs with the result that the statement refers back in time, such that it is equivalent to a single sentence version of the Liar's Paradox in (\ref{liar}). 
 }
\label{Circle}\end{figure}

\section{Self-Reference and Cyclical Loop}
In quantum theory, there are two main variables: the state vector, which provides a mathematical description of the object being observed; and an observable, a mathematical description of the reference frame of the observing party. These two descriptions are exact rather than an approximation. With the description of the observing party and the object, the process of self-observation corresponds with that of the object being observed is the observer, or the reference frame of the observer. If we denote the reference frame of the observer as RF, the physical version of the Liar's Paradox may be written as follows:  
\begin{equation}
An \; observer\; in\; RF\; observes\; the\; observer's\; own\; RF. 
\label{self}\end{equation}
Just as self-reference leads to inconsistency in a logical structure, the self-observation shown in (\ref{self}) leads to a non-computable aspect among the natural phenomena that are ordinarily computable \cite{song2007}.

There is another way of presenting the Liar's Paradox in (\ref{liar}), which is using the following two sentences: 
\begin{enumerate}[label=(\roman*)]
\item  The following sentence is true
\item  The previous sentence was not true
\end{enumerate}
The two-sentence structure of the Liar's Paradox has the same paradoxical outcome as in  (\ref{liar}). However, it removes the ambiguity of {\it{This sentence}} in (\ref{liar}), which refers to a yet undefined whole sentence at the expense of adopting a cyclical loop, also known as a strange loop \cite{hofstadter}.

As pointed out by several authors (for example, see \cite{hofstadter,reynolds}),  the cyclical loop above contains a causal or time factor. That is, initially, the two-sentence Liar's Paradox is evolving forward in time, while (ii) refers to the past, or travels backward in time (Fig. \ref{Circle}). Following this analogy of two-sentence Liar's Paradox, the physical version of the self-reference in (\ref{self}) may also be written in two-sentence structure as follows (also see \cite{song2017b}): 
\begin{enumerate}[label=(\roman*)]
\item An observer in RF makes an observation 
\item The observed was the observer's RF 
\end{enumerate}
That is, the observer observes in a time-forward manner, yet the observed object turns out to be the observer's own reference frame in a time-backward manner. In the following, we will argue that this cyclical loop property of consciousness may help to resolve the cosmological constant problem.

\begin{figure}
\begin{center}
{\includegraphics[scale=.65]{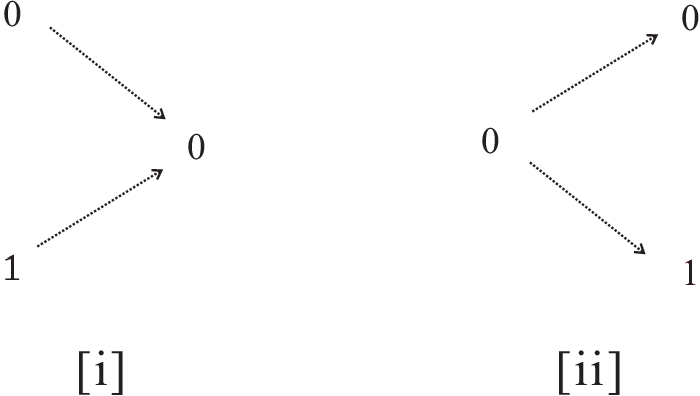}}

\end{center}
\caption{ Landauer has shown that the irreversible computation, or the information erasure process, as in [i] yields an energy dissipation of $kT\ln 2$ and an entropy of $k\ln 2$. The nondeterministic computation [ii] corresponds to the time-reversal process of [i].
 }
\label{DiracSea}\end{figure}

\section{Cyclical Looping}
Over the years, studies have suggested a deep connection between the physical system and computation (for example, see \cite{deutsch,landauer}).  For instance, Wheeler coined the term {\it{it from bit}} \cite{wheeler}, suggesting information plays a fundamental role in the description of natural phenomena. Others have suggested that we view the whole universe in terms of information and computation (for example, see \cite{thooft,wolfram,schmidhuber,lloyd2006,vedral,zizzi}). In \cite{lloyd}, based on the energy involved in elementary computations, the total number of computations of the observable universe since the big bang has been estimated, which is
\begin{equation}
N \sim  10^{123}
\label{total}\end{equation}
The number of elementary computations in (\ref{total})  roughly coincides with the discrepancy between the observation and the theoretical values of vacuum energy in (\ref{discre})  (for example, see \cite{funkhouser}).

In \cite{landauer}, Landauer has discussed the irreversible process (i.e., both the initial state 0 or 1 is set to 0 (Fig. \ref{DiracSea}  [i])) that necessarily dissipates energy of $kT\ln 2$ (this result has also been useful in resolving Maxwell's demon problem \cite{bennett}). Moreover, one may consider the time reversal process of irreversible computation, namely, non-deterministic computation, which chooses an acceptable path as in Fig. \ref{DiracSea} [ii].  In \cite{song2014b}, we have argued that the physical process of nondeterministic computation occurs with decision-making phenomenon in self-observing consciousness, such that it is not computable by ordinary computational processes. This provides a physical example why a deterministic computer in polynomial time, or P, cannot compute a non-deterministic process in polynomial time, or NP.

As discussed earlier, there are two main variables in quantum theory. Moreover, in the process of quantum measurement, there are additional aspects involved. While state vectors and observables are defined in complex vector space, the actual measurement takes place in classical space, i.e., both the object and the reference frame of the observer are defined not only in quantum Hilbert space but in classical space.  Since we are concerned with the case in which the object being observed is the observer's reference frame, only the classical and quantum reference frames will be considered.

\begin{figure}
\begin{center}
{\includegraphics[scale=.7]{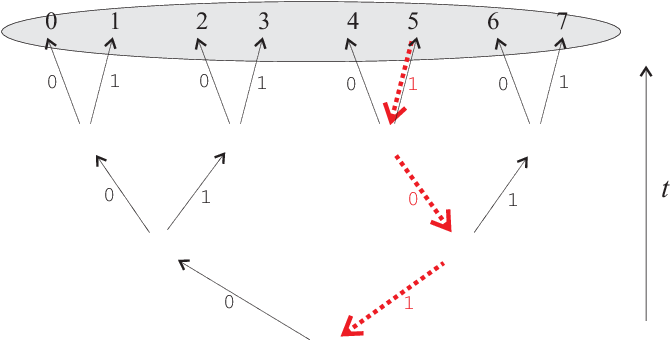}}

\end{center}
\caption{A toy model may be considered with three irreversible computations. In a time-forward manner, a single irreversible computation is performed. The cyclical loop is completed with the choice following one of eight possible paths in time going backwards. 
 }
\label{ToyModel}\end{figure}

\section{The Decision-Making Model}
As shown in Figure \ref{ToyModel}, let us consider a toy model with three computational processes where each is an irreversible process, erasing one bit of information and therefore giving $k\ln2$ of entropy. The computation will be considered as the evolution of the observer's reference frame. The cyclical process of consciousness discussed previously in terms of the observer's reference frames may be argued as follows: In the first part of the loop, the observer's classical reference frame evolves, i.e., an irreversible computation in a time-forward manner. The second part of the loop, i.e., the one that closes the loop, which travels backward in time, the choice may become one of the eight paths, i.e., the evolution of the quantum reference frame. This may be summarized as:
\begin{enumerate}[label=(\roman*)]
\item  An observer's classical reference frame evolves
\item  The quantum reference frame follows one of the eight possible paths through three nondeterministic computations.
\end{enumerate}
The above may be understood in the following sense: The observer's reference frame evolves in classical or physical space forward in time, yet the conscious realization of the choice is done in a time-backward manner, or through nondeterministic computation, such that the whole cyclical process is equivalent to the single-sentence version of consciousness as in (\ref{self}). If we assume the same amount of energy is generated in each elementary computation, the energy generated in the time-forward manner is $E_0$ while the energy corresponding to the case of the second part of the cyclical loop corresponds to $3E_0$.

Let us now apply the same logic to the computational model of the universe with $N$ number of computations in (\ref{total}). Like the toy model in Fig. \ref{ToyModel}, it will be assumed that there are two vectors, rather than one, that correspond to the observer's reference frame where one is classical and the other is quantum. With this assumption, one may consider the following process:

\begin{enumerate}[label=(\roman*)]
\item  The observer's classical reference frame evolves 
\item   The quantum reference frame evolves through $N$ number of nondeterministic computations.
\end{enumerate}

\begin{figure}
\begin{center}
{\includegraphics[scale=.55]{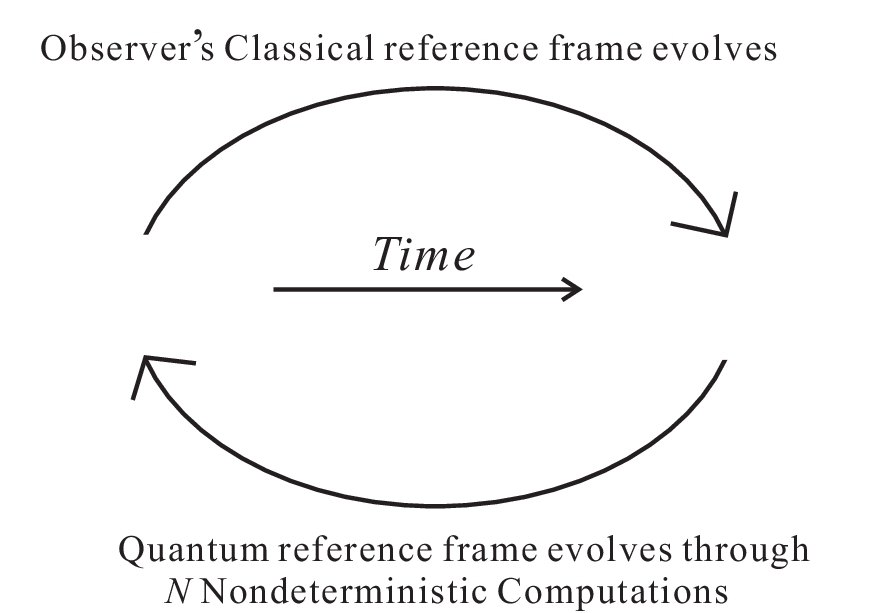}}

\end{center}
\caption{ Physical version of the two-sentence Liar's Paradox. In a time-forward manner, the classical reference frame in which observer exists evolves. Then the cyclical looping occurs, such that the quantum reference frame evolves in a time-backward manner. 
 }
\label{Physical}\end{figure}

While the classical vector chooses, i.e., one unit of classical computation is carried out in the time-forward manner as in (i), the quantum reference frame evolves backward in time to not only a single unit of computation, but all the way back to the big bang with $N$ number of computations (Fig. \ref{Physical}). The energy density in the process of classical computation should correspond to the classical vacuum, i.e., $\rho_{\Lambda}$, while the energy density corresponding to the quantum evolution that goes backwards in time would be classical density multiplied by the number of computations, i.e.,
\begin{equation}
 N \cdot \rho_{\Lambda}
\label{relationi}\end{equation}
Previously, the vacuum energy predicted from quantum theory was thought to correspond to the observed value of the vacuum energy, which resulted in one of the largest discrepancies between theory and experimental verification. However, as argued above, the energy calculated from quantum theory should correspond to the energy of the negative sea that fills up the universe as the conscious reference frame of the observer.

\section{Discussion}
For centuries, science has been studied on the basis of causality. That is, for a given result, there is a corresponding cause. This method of revealing an objective pattern in natural phenomenon began to shatter with the development of quantum mechanics at the beginning of the twentieth century. Indeed, rather than the objective description for a given physical system, quantum theory started to describe the subjective relation between the observer and the object being observed. 

In \cite{song2012}, this subjective description, which was often considered an incomplete aspect of a full description  \cite{EPR}, may correspond to not only the limit of scientific knowledge but existence itself. By considering the observable in quantum theory as the reference frame of the observing party, the two-picture formulation of quantum theory no longer correctly described self-observing consciousness; the observer and the object being observed (i.e., the universe) should be inseparable. In \cite{song2016}, a more specific description of what it means for the observer and the universe to be inseparable has been provided. The nature of negative sea was discussed in more detail by identifying it as the time reversal of the irreversible computation since the big bang. 

In this paper, we have discussed the difference between the observed and the theoretical values regarding vacuum energy. We have used the cyclical looping in time to discuss the vacuum as the physical realization of consciousness. That is, a vacuum should correspond to the unit of classical computation of the observer's reference frame, which is filled with the negative energy of the time-reversal process of irreversible quantum computation. The result obtained in this paper indeed strengthens the ongoing efforts in building the subjective model where the discrete and finite classical vacuum is filled with a continuous and infinite Dirac-type negative sea of consciousness \cite{song2017a}.

%%%%%%%%%%%%%%%%%%%%%%%%%%%%%%%%%%%%%%%%%%%%%%%%%%%%%%%%%%%%%%%%%%%%%%%%
%%%%%%%%%%%%%%%%%%%%%%%%%%%%%%%%%%%%%%%%%%%%%%%%%%%%%%%%%%%%%%%%%%%%%%%%
%%%%%%%%%%%%%%%%%%%%%%%%%%%%%%%%%%%%%%%%%%%%%%%%%%%%%%%%%%%%%%%%%%%%%%%%
%%%%%%%%%%%%%%%%%%%%%%%%%%%%%%%%%%%%%%%%%%%%%%%%%%%%%%%%%%%%%%%%%%%%%%%%
%%%%%%%%%%%%%%%%%%%%%%%%%%%%%%%%%%%%%%%%%%%%%%%%%%%%%%%%%%%%%%%%%%%%%%%%

\end{document}